# Multidimensional Replica Exchange simulations for Efficient constant pH and Redox Potential Molecular Dynamics

*Vinícius Wilian D. Cruzeiro [1], Adrian E. Roitberg [1,\*]*

[1] Department of Chemistry, University of Florida, Gainesville, FL 32611, United States

ABSTRACT

Efficient computational methods that are capable of supporting experimental measures obtained at constant values of pH and redox potential are important tools as they serve to, among other things, provide additional atomic level information that cannot be obtained experimentally. Replica Exchange is an enhanced sampling technique that allows converged results to be obtained faster in comparison to regular molecular dynamics simulations. In this work we report the implementation, also available with GPU-accelerated code, of pH and redox potential (E) as options for multidimensional REMD simulations in AMBER. Previous publications have only reported multidimensional REMD simulations with the temperature and Hamiltonian dimensions. In this work results are shown for N-acetylmicroperoxidase-8 (NAcMP8) axially connected to a histidine peptide. This is a small system that contains only a single heme group. We compare results from E,pH-REMD, E,T-REMD and E,T,pH-REMD to one dimensional REMD simulations and to simulations without REMD. We show that two-dimensional REMD simulations improve sampling convergence in comparison to one-dimensional REMD simulations, and that three-dimensional REMD further improves convergence in comparison to two-dimensional REMD simulations. Also, our computational benchmarks show that our multidimensional REMD calculations have a small and bearable computational performance, essentially the same as one dimensional REMD. However, in multidimensional REMD a significantly higher number of replicas is required as the number of replicas scales geometrically with the number of dimensions, which requires additional computational resources. In addition to the pH dependence on standard redox potential values and the redox potential dependence on p$K_a$ values, we also investigate the influence of the temperature in our results. We observe an agreement between our computational results and theoretical predictions.

INTRODUCTION

Many biochemical studies are conducted at both constant pH and constant redox potential. The charge distribution in molecules is affected by the solvent pH and redox potential due to changes in the protonation/redox state of relevant groups. These states can be related to the structure and function of proteins and other biomolecules, affecting properties like stability, ligand binding, catalysis, absorption spectrum, among others [1,2]. Theoretical methods that can correctly sample protonation/redox states at constant pH and redox potential can then have an important role in the study of many systems. Constant pH simulations have been extensively employed in the study of different systems [3–9]. Some publications also reported simulations at both constant pH and redox potential [10–17]. Recently, we have implemented the constant pH and redox potential MD (C(pH,E)MD) method in AMBER [18]. With this implementation it is now possible in AMBER to predict the standard Redox Potential ($E^o$) of redox-active titratable residues, like heme groups, with the same easiness and accuracy as the $pK_a$ of pH-active residues obtained through constant pH MD simulations. The C(pH,E)MD method is also available using AMBER's GPU-accelerated code. The GPU-accelerated C(pH,E)MD simulations have a high computational performance, which represents a breakthrough in comparison to previous CPU-based software packages that allow the study of redox potential effects.

Replica Exchange Molecular Dynamics (REMD) is a state-of-the-art technique that allows faster statistical ensemble convergence while taking advantage of computational parallelism. This is achieved by periodically exchanging information between neighbors (independent replicas) throughout the simulation using Metropolis Monte Carlo exchange attempts. Simulations without replica exchange, mainly for large systems, may require a large number of steps to achieve ensemble convergence. By exchanging information between independent replicas, REMD simulations have shown to significantly improve statistical sampling [4,9,19–23].

Previous studies have shown that, in some cases, increasing the Exchange Attempt Frequency (EAF) in replica exchange simulations may lead to a better sampling convergence [4,19,24]. Sugita *et. al.* [25] and Bergonzo *et. al.* [26] have shown that combining Hamiltonian and temperature dimensions in replica exchange simulations (H,T-REMD) further improves sampling in comparison to both T-REMD and H-REMD. In this work, we put this concept into test for multidimensional simulations along the pH, redox potential and temperature dimensions. We have recently implemented the pH and redox potential dimensions into the multidimensional REMD module in AMBER [27]. This implementation represents additional options to the existing temperature and Hamiltonian dimensions, it is part of the recent AMBER 18 release [27], and is available in both *sander* and *pmemd* including using AMBER's GPU-accelerated code. To the best of our knowledge, this is the first work to report pH or redox potential as part of multidimensional replica exchange simulations.

In this work, we are going to present results for N-acetylmicroperoxidase-8 (NAcMP8) [28] with a histidine peptide as the axial ligand. This system is derived from cytochrome *c* and has a single redox-active heme group. Results will be presented and evaluated for long and short runs of multidimensional REMD, one dimensional REMD and simulations without replica exchange in order to study the sampling efficiency of each type of simulation.

## THEORY, METHODS AND IMPLEMENTATION

**Constant pH and Redox Potential Molecular Dynamics (C(pH,E)MD)**

Our C(pH,E)MD approach makes use of Monte Carlo transitions between discrete protonation or redox states, represented by different atomic charge distributions for protonated and deprotonated states of a given pH-active residue and reduced and oxidized states of a given redox-active residue. In our approach, MD is performed for a predetermined number of steps and then the simulation is halted. Afterward, a protonation and/or a redox state change is attempted. Protonation and redox state change attempts are performed separately. The interval between state change attempts is tunable and may be different for protonation and redox state change attempts. If a protonation and a redox state change attempts are to be performed at the same MD step, the protonation state change attempt is executed first. The relation between pH and redox potential effects comes naturally in our methodology because the charge distribution change in a pH-active residue that had a successful protonation state change attempt will affect the following redox state change attempts in neighboring redox-active residues. In the same way, a successful redox state change will affect the protonation state change attempts of pH-active groups nearby. More details about the C(pH,E)MD method and its implementation in AMBER can be found in our previous publication [18].

The fraction of protonated species $f_{prot}$ of a pH-active residue and the fraction of reduced species $f_{red}$ of a redox-active residue can be described, respectively, as a function of pH and redox potential using the following equations:

$$f_{prot} = \frac{1}{1 + e^{n_{prot}(\ln 10)(pH - pK_a)}} \qquad (1)$$

$$f_{red} = \frac{1}{1 + e^{n_{red}\frac{vF}{k_bT}(E - E^o)}} \qquad (2)$$

In these equations, $n_{prot}$ and $n_{red}$ are Hill coefficients, $v$ is the number of electrons involved in the reduction reaction, $F$ is the Faraday constant, and $E$ and $E^o$ are, respectively, the redox potential and the standard redox potential, the equivalent of $pH$ and $pK_a$ in electrochemistry.

**Multidimensional Replica Exchange MD along the pH, redox potential and temperature dimensions**

*Replica Exchange along the pH dimension*

In pH-REMD, each replica explores the conformational space at a different value of pH. The target pH is the information exchanged periodically across replicas throughout the simulation. The Metropolis Monte Carlo probability of exchanging any two replicas in pH-REMD is [4,29]:

$$P_{i \to j} = \min\left\{1, e^{\ln(10)(pH_i - pH_j)\left(N_i^{H^+} - N_j^{H^+}\right)}\right\} \qquad (3)$$

where $pH_i$ and $N_i^{H^+}$ are, respectively, the pH and the number of protons in replica $i$. As can be seen, this probability only depends on the difference of pHs and number of protons between the replicas involved.

*Replica Exchange along the redox potential dimension*

In E-REMD, each replica has a different value of redox potential, and these are the values exchanged throughout the simulation. In E-REMD the probability of exchanging any two replicas is [18]:

$$P_{i \to j} = \min\left\{1, e^{\frac{F}{k_bT}(E_i-E_j)\left(N_i^{e^-}-N_j^{e^-}\right)}\right\} \qquad (4)$$

where $F$ is the Faraday constant, and $E_i$ and $N_i^{e^-}$ are, respectively, the redox potential and the number of electrons in replica $i$. Similarly to pH-REMD, in E-REMD the probability of exchange only depends on the difference of redox potential and number of electrons of the replicas that are being exchanged.

*Replica Exchange along the temperature dimension*

In T-REMD, each replica explores the conformational space at a different target temperature and has its target temperature exchanged periodically throughout the simulation. The probability of exchanging any two replicas is given by [19]:

$$P_{i \to j} = \min\left\{1, e^{\left(1/k_bT_i - 1/k_bT_j\right)(\varepsilon_i-\varepsilon_j)}\right\} \qquad (5)$$

where $T_i$ and $\varepsilon_i$ are, respectively, the temperature and the energy of the system in replica $i$. As the equation shows, the probability of exchange in T-REMD is not only related to the temperature difference between the replicas being exchanged but it also depends on the energy difference between the replicas. Because of the energy difference dependence, oppositely to pH-REMD and E-REMD, in T-REMD the distribution of target temperatures to cover a given temperature range with a good acceptance rate varies from system to system. It can be shown that the number of replicas required to cover well a given temperature range increases with the square root of the number of degrees of freedom in the system [30,31]. Because of this, explicit solvent simulations generally require a significantly higher number of replicas in comparison to implicit solvent simulations to cover the same temperature range.

*AMBER's multidimensional REMD module*

It is possible to devise expressions for the probability of exchange when more than one type of information is exchanged at the same time. Exchanging more than one type of information at the same time may either increase or decrease the exchange probability. Therefore, one strategy is to exchange only one type of information at a time. Also, the exchange of multiple information at the same time would require a different expression for each multidimensional exchange scheme, whereas by exchanging one type of information at a time the exchange probability expressions reduce to the one-dimensional REMD expressions (as in equations 3, 4 and 5), which also simplifies the computational implementation of multidimensional REMD. For this reason, in AMBER's multidimensional REMD module only one type of information can be exchanged each time a replica exchange is to be attempted.

In AMBER, the number of MD steps between two consecutive exchange attempts is determined by the Exchange Attempt Frequency (EAF), which is a tunable variable to be input by the user. The order of dimensions to be exchanged, which exemplify in whether doing E,pH,T-REMD or T,pH,E-REMD, can also be determined by the user. In E,pH,T-REMD, for example, the first exchange attempt would be in the redox potential dimension, the second in the pH dimension, and the third in the temperature dimension. This sequence would then be repeated throughout the whole simulation. Also, redox potential exchange attempts would only be performed between replicas with the same pH and temperature. Similarly, pH exchange attempts would only happen between replicas with the same redox potential and temperature, and temperature exchange attempts would only occur between replicas with the same redox potential and pH. Assuming a distribution of redox potential, pH and temperature values across the replicas which would provide a good exchange acceptance rate in all dimensions, this scheme allows a given replica to visit all possible target redox potential, pH and temperature values.

It is important to mention that equations 3, 4 and 5 show that the probability of acceptance is higher between replicas that have, respectively, pH, redox potential, and temperature close to each other. For this reason, the one-dimensional exchange attempts happen between nearest neighbors.

Additional information about AMBER's multidimensional REMD module can be found in the publication by Bergonzo et al. [26] and in AMBER 18's user manual [27].

**Describing theoretically the pH-dependence of $E^o$ and the redox potential-dependence of $pK_a$ values**

Assuming the system contains only a single redox-active residue and one or more pH-active residues, by making thermodynamic considerations it is possible to show that [18]:

$$vF\left(E^o_{prot} - E^o_{deprot}\right) = k_bT \ln(10) \sum_i \left(pK^{(i)}_{a,red} - pK^{(i)}_{a,oxi}\right) \quad (6)$$

$$E^o = E^o_{prot} + \frac{k_bT}{vF} \sum_i \ln\left(\frac{10^{-pK^{(i)}_{a,red}} + 10^{-pH}}{10^{-pK^{(i)}_{a,oxi}} + 10^{-pH}}\right) \quad (7)$$

$$\sum_i pK^{(i)}_a = \sum_i pK^{(i)}_{a,red} + \log\left(\frac{e^{-vFE^o_{prot}/k_bT} + e^{-FE/k_bT}}{e^{-vFE^o_{deprot}/k_bT} + e^{-FE/k_bT}}\right) \quad (8)$$

where $v$ is the number of electrons involved in the reduction reaction, $E^o_{prot}$ and $E^o_{deprot}$ are the standard redox potential of the redox-active residue when all the pH-active residues are, respectively, fully protonated or fully deprotonated, and $pK^{(i)}_{a,red}$ and $pK^{(i)}_{a,oxi}$ are the $pK_a$ values of the $i^{th}$ pH-active residue when the redox-active residue is, respectively, fully reduced or fully oxidized. The derivations of all these equations can be found at the Supporting Information of our previous publication [18].

From equation 6 we can see that the difference between $E^o_{prot}$ and $E^o_{deprot}$ is directly related to differences between $pK^{(i)}_{a,red}$ and $pK^{(i)}_{a,oxi}$. $E^o$ is the pH-dependent standard redox potential value of the redox-active residue when the pH-active residues are not all fully protonated nor all fully deprotonated. In the low pH limit, $E^o$ becomes $E^o_{prot}$. In the high pH limit, $E^o$ becomes $E^o_{deprot}$ and equation 7 turns into equation 6. $pK^{(i)}_a$ is the redox potential-dependent $pK_a$ value of the $i^{th}$ pH-active residue. Similarly to equation 7, in the limit of high redox potential values $pK^{(i)}_a$ becomes $pK^{(i)}_{a,oxi}$ and equation 8 becomes equation 6.

CALCULATION DETAILS

We performed all the simulations in this work using AMBER 18 [27]. AMBER 18 contains new tools we implemented aiming at making it easier to work with multidimensional REMD simulations. The *genremdinputs.py* tool allows users to automatically create all input files for any replica exchange simulation, with any number of dimensions and exchanging dimensions in any order the user desires. With *fixremdcouts.py* users are able to reorder for posterior analyzes the constant pH and/or constant redox potential output files obtained with any one-dimensional or multidimensional REMD simulation.

AMBER allows multidimensional REMD simulations to be performed in both implicit and explicit solvent models. For simplicity, in this work we will be only performing implicit solvent calculations. Below in this section, we are going to discuss details about the parametrization of our system, followed by details of the simulations performed.

**Parametrization of NAcMP8 with a histidine peptide**

Figure 1 shows N-acetylmicroperoxidase-8 (NAcMP8) axially connected to a histidine peptide. On the other side of the porphyrin plane, NAcMP8 has a histidine residue from its peptide chain that is also axially connected to the heme group. The E° of NAcMP8 axially connected to an imidazole molecule, a situation similar to the system shown in Figure 1, has been experimentally measured as -203 mV *vs.* NHE at pH 7.0 [28].

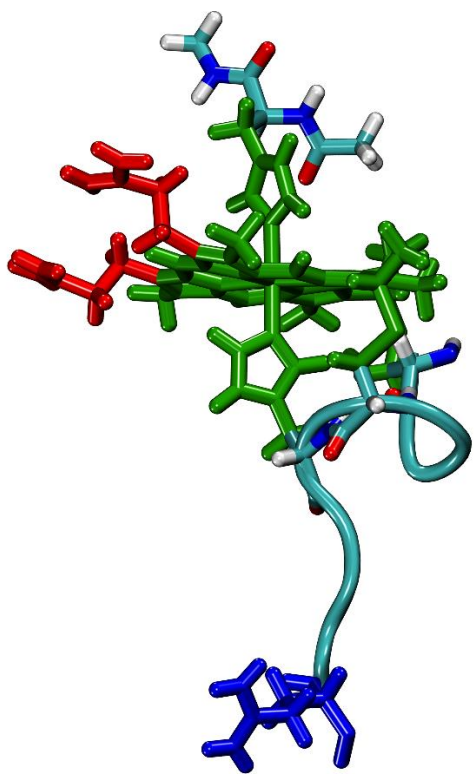

**Figure 1.** N-acetylmicroperoxidase-8 (NAcMP8) axially connected to a histidine peptide. The redox-active residue HEH is shown in green. The two pH-active propionates (PRN) are shown in red. The pH-active glutamate (GL4) is shown in blue.

In our parametrization, the redox-active residue that changes its atomic charge distribution when a redox state change attempt is accepted is called HEH. The HEH residue is shown in green at Figure 1 and is composed by the ferric porphyrin ring together with the side chains of two histidines and two cysteines. The two heme propionates (PRN) are colored in red in Figure 1, are pH-active residues, and are separate residues from HEH. Our system also contains one pH-active glutamate (GL4), colored in blue in Figure 1.

Our force field parameters are available in AMBER's *tleap* module through the *leaprc.conste* repository. The parameters for the heme group were taken from Crespo *et al.* [32], excluding the atomic charges. The charges for both the reduced and oxidized states were adapted from Henriques *et al.* [33]. All other residues have parameters from the AMBER FF99SB force field [34].

The reference energies necessary for C(pH,E)MD that we used are available in AMBER's *ceinutil.py* tool [18] for the residue HEH and in the *cpinutil.py* tool [4] for the residues PRN and GL4. These reference energies are computed in such a way that simulations of the reference compounds yield the experimental $E^o$ or $pK_a$ values. For PRN residues the experimental $pK_a$ of a propionic acid in water is taken as the reference, therefore $pK_{a,ref} = 4.85$ [35]. For the HEH residue the reference energy is taken in such a way that by performing C(pH,E)MD simulations for our system, NAcMP8 with a histidine peptide as the axial ligand, we obtain the experimental standard redox potential at pH 7.0, thus $E^o_{ref} = -203$ mV. For the GL4 residue the reference compound is a glutamate in water and we have $pK_{a,ref} = 4.4$.

**Implicit Solvent Calculations**

We start with the heme group in the oxidized state, and with the glutamate and the propionates in the deprotonated state. A minimization is then performed for 100 steps using the steepest descent algorithm and then for 3900 steps using the conjugate gradient algorithm constraining the backbone atoms with a 10 kcal/mol·Å² constant. Afterward, the minimized structure is heated for 3 ns. During heating, the temperature is controlled using Langevin dynamics with a friction frequency of 5 ps$^{-1}$, the target temperature is varied linearly from 10 to 300 K over the initial 0.6 ns, and the backbone atoms are constrained using a force constant of 1 kcal/mol·Å$^2$.

After heating, we perform an equilibration simulation at 300 K during 10 ns using Langevin dynamics with a 10 ps$^{-1}$ friction frequency and a 0.1 kcal/mol·Å$^2$ backbone constraint. The equilibrated structure is then used as the initial structure in our production simulations. In the production simulations, no positional restraints are applied, the friction frequency of the Langevin dynamics is 10 ps$^{-1}$, and redox and protonation state change attempts are performed every 10 fs. All REMD calculations were performed using an Exchange Attempt Frequency (EAF) of 25 ps$^{-1}$, which means one exchange attempt every 40 fs. The time step used in all simulations is of 2 fs, and the lengths of all hydrogen bonds were constrained using the SHAKE algorithm [36,37].

Following previous AMBER implicit solvent CpHMD publications [3,4,38] and our previous C(pH,E)MD publication [18], in order to account for solvent effects we have used the Generalized Born model proposed by Onufriev *et al.* [39] (*igb*=2 in AMBER). This model is used during MD and also during protonation and redox state change attempts.

RESULTS AND DISCUSSIONS

**Evaluating two-dimensional REMD simulations: long runs**

In this subsection, we will evaluate how the addition of a second replica exchange dimension affects sampling convergence in comparison to one-dimensional REMD simulations and simulations without replica exchange. This will be done by analyzing how our standard redox potential predictions behave as a function of simulation time for long runs.

*E,pH-REMD at 300 K*

Here, our production simulations were executed for 110 ns. Simulations were performed using a target temperature of 300 K, 6 redox potential values from -263 to -113 mV in intervals of 30 mV, and 6 pH values from 5.0 to 7.5 in intervals of 0.5, giving a total of 36 replicas. The cumulative fraction of reduced species for each combination of redox potential and pH values was obtained as a function of time. For each time we gathered the fractions for all redox potential values at a given pH to fit $E^o$ using equation 2. By doing this, the cumulative prediction of $E^o$ as a function of simulation time for each pH value can be obtained. We repeated this procedure independently 5 times. In Figure 2 we show the average for the 5 independent simulations of the cumulative $E^o$ as a function of simulation time for each pH. The standard deviation of the $E^o$ predictions in the 5 independent simulations as a function of time is also shown in the figure for each pH. Results are presented for E,pH-REMD, E-REMD, pH-REMD and C(pH,E)MD. C(pH,E)MD corresponds to the limit of EAF=0, which means no replica exchange attempts were performed. In E-REMD, replicas are only allowed to have their redox potential values exchanged, and equivalently in pH-REMD only pH values can be exchanged.

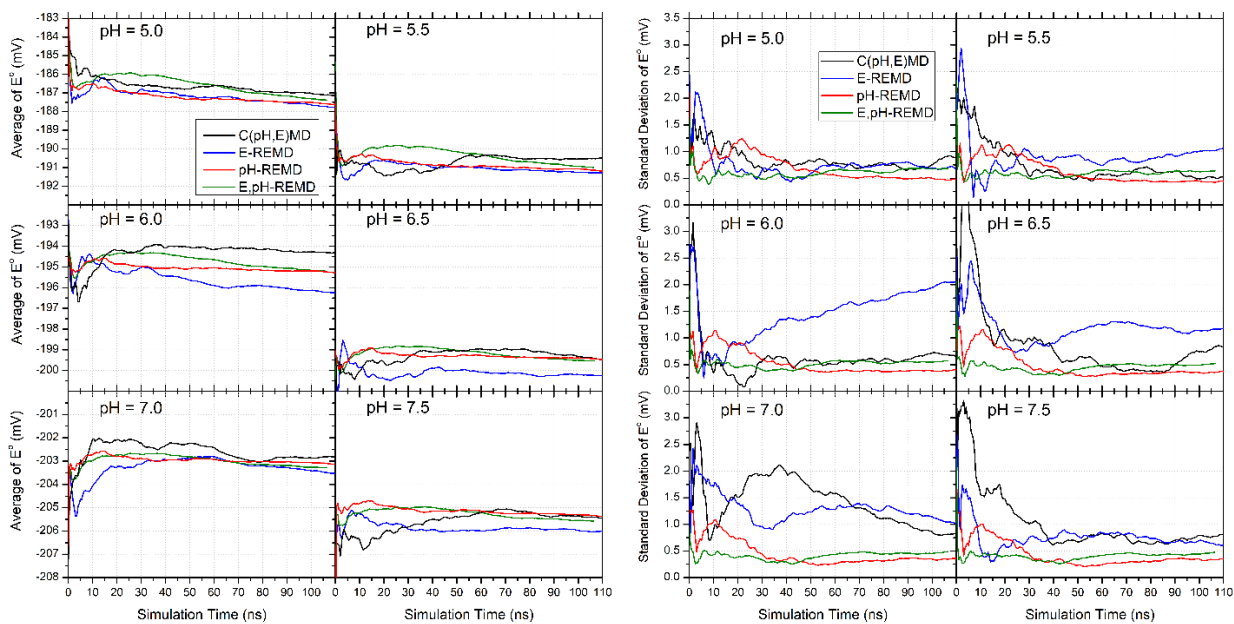

**Figure 2.** Cumulative standard redox potential (E°) of HEH averaged for 5 independent simulations (left) and its standard deviation (right) as a function of time for different pH values. Simulations were performed with a target temperature of 300 K.

As Figure 2 shows, for all simulations performed the values of E° decrease when pH increases. This is expected because the concentration of deprotonated species raises as the pH increases, which makes it less energetically favorable for an electron to reduce a redox-active group, causing then lower E° values.

If we interpret the E° standard deviations as a measure of error bars for our methodologies, we see that the average E° predictions from C(pH,E)MD, E-REMD, pH-REMD and E,pH-REMD agree with each other within error bars.

However, in the figure we see that the standard deviations for E,pH-REMD are small, even for short simulation times, and remain small throughout the whole simulation, fluctuating significantly less than the standard deviations for all other simulations. This can be concluded for all pH values. Lower errors for E,pH-REMD in comparison to the other simulations indicate a better sampling efficiency. If we look at the results at 20 ns, for example, we see that E,pH-REMD has smaller errors than both E-REMD and pH-REMD. By comparing E-REMD and pH-REMD at 20 ns for the 6 pH values simulated, we see that for 2 pH values (6.0 and 6.5) the errors basically the same for E-REMD and pH-REMD, for 3 pH values (5.0, 5.5 and 7.5) the errors for E-REMD are lower, and for 1 pH value (7.0) the error is smaller for pH-REMD. For longer simulation times these trends change and the errors for pH-REMD become comparable to E,pH-REMD and lower than E-REMD, excluding for pH 5.0 where pH-REMD, E-REMD and even C(pH,E)MD have similar errors to E,pH-REMD for long simulation times.

It draws attention to notice, around 25 ns for E-REMD at pH 6.0, the sudden decrease in $E^o$ accompanied by an increase in the standard deviation. In our previous publication [18], we have identified an infrequent conformational change that is responsible for dropping the $E^o$ prediction. This conformational change is related to a flip in the histidine axially connected to the heme group that is part of the NAcMP8 peptide chain. We verified that this conformational change happened around 25 ns in one our E-REMD independent simulations at pH 6.0. The fact that this is happening in only one of the independent simulations explains the sudden increase in the standard deviation values observed for E-REMD at pH 6.0.

*E,T-REMD at pH 7.0*

In this analysis our production simulations were executed for 140 ns. We performed simulations at pH 7.0 for 6 redox potential values from -263 to -113 mV in intervals of 30 mV, and 6 target temperatures from 280 to 380 K in intervals of 20 K. This gives a total of 36 replicas considered. The distribution of temperatures we used was selected with the aid of a model [31], aiming at obtaining a good exchange acceptance rate in the temperature dimension for our system.

Similarly to the procedure described in the previous subsection, the cumulative prediction of E° was obtained as a function of simulation time for each target temperature. Figure 3 contains the cumulative E° averaged for 5 independent simulations as a function of simulation time for each target temperature. Standard deviations are also shown. We present results for E,T-REMD, E-REMD, T-REMD and C(pH,E)MD.

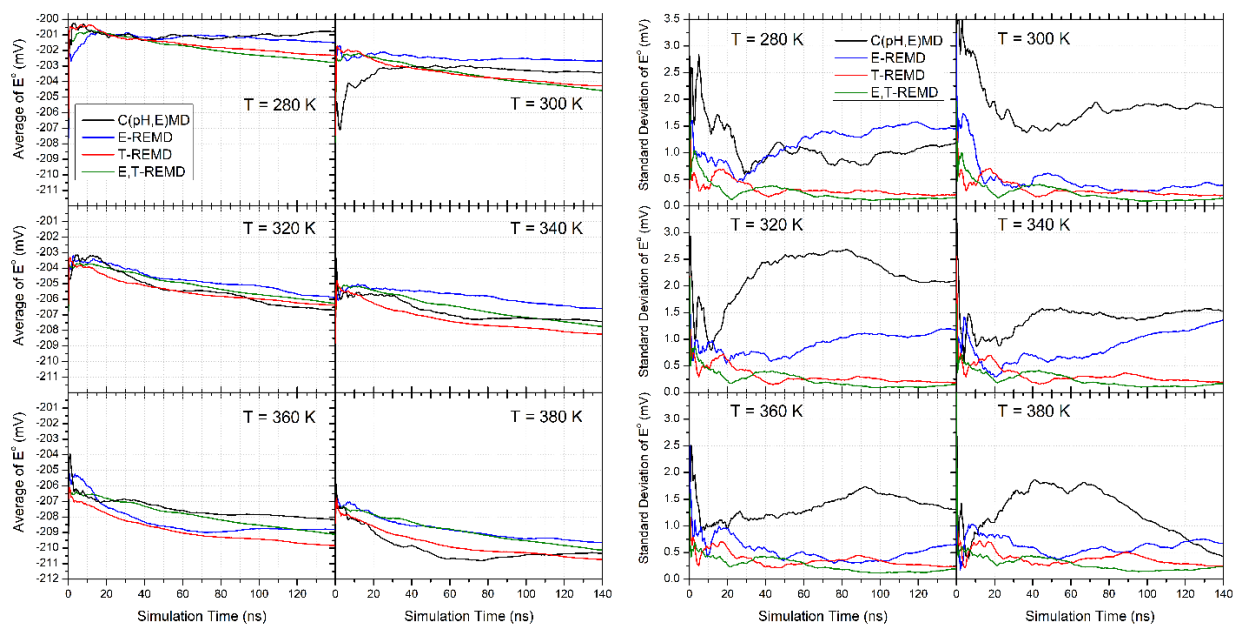

**Figure 3.** Cumulative standard redox potential (E°) of HEH averaged for 5 independent simulations (left) and its standard deviation (right) as a function of time for different target temperatures. Simulations were performed at pH 7.0.

As can be seen from Figure 3, increasing the temperature leads to a decrease in the E° values. Similarly to E,pH-REMD in Figure 2, we observe low errors for E,T-REMD in comparison to all other simulations even for small simulation times. This means E,T-REMD has a better sampling efficiency. At 20 ns, in all temperatures simulated the errors for E,T-REMD are smaller than for all the other simulations. By comparing E-REMD and T-REMD at 20 ns, we see the errors are essentially the same in 1 temperature (320 K), are lower for E-REMD in 2 temperatures (300 and 340 K), and are lower for T-REMD in 3 temperatures (280, 360 and 380 K). For larger simulation times the errors for T-REMD get close to the errors for E,T-REMD and significantly smaller than E-REMD.

*E,pH-REMD vs E,T-REMD at 300 K and pH 7.0*

By evaluating the data at 300 K and pH 7.0, common to both E,pH-REMD and E,T-REMD, we can infer what dimension, pH or temperature, gives better sampling convergence when added to E-REMD. For an easier comparison, in Figure 4 we extracted the average E° and its standard deviation at 300 K and pH 7.0 for E,pH-REMD and E,T-REMD from, respectively, Figures 2 and 3.

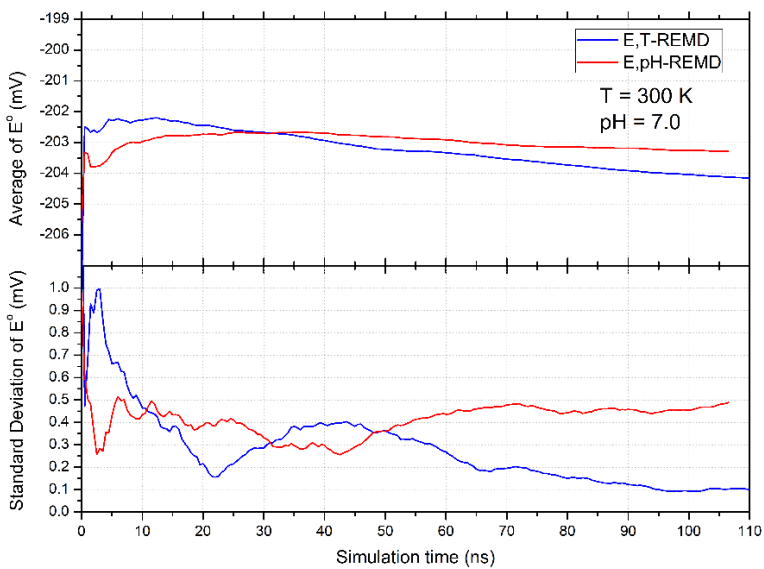

**Figure 4.** Comparison of average E° and its standard deviation between E,pH-REMD and E,T-REMD at 300 K and pH 7.0.

As Figure 4 shows, the standard deviations are smaller for E,T-REMD. For large simulation times, we see the E,T-REMD error is significantly smaller than for E,pH-REMD. This indicates that for an improved convergence efficiency the addition of a temperature dimension to E-REMD is better than the addition of a pH dimension. This conclusion can be reasoned by the fact that with the addition of a temperature dimension the system can visit higher temperatures which gives more mobility to better explore the conformational space.

**Evaluating three-dimensional REMD simulations: short runs**

We now evaluate the effect of adding a third REMD dimension to the simulations. We will do that here by presenting results for short simulations where C(pH,E)MD is still not converged. As replica exchange simulations are meant to accelerate convergence, we then evaluate the effect of performing different replica exchange calculations by comparing E,T,pH-REMD, E,pH-REMD, E,T-REMD and C(pH,E)MD results for the same number of total MD steps.

In this analysis, all simulations were executed for 5 ns. Simulations were performed for 6 redox potential values equally spaced from -263 to -113 mV, for 6 temperatures from 280 to 380 K equally spaced, and for 6 pH values equally spaced from 5.0 to 7.5, totalizing 216 replicas. By doing fittings with equation 2 we obtained the values of the E° of HEH from the fractions of reduced species for all redox potential values for each combination of temperature and pH. Similarly, the p$K_a$ of one of the propionates (PRN 16) and of the glutamate (GL4 12) were obtained using equation 1 and the fractions of protonated species for all pH values for each combination of temperature and redox potential. With this, we are able to plot E° as a function of temperature and pH, and the p$K_a$ of PRN 16 and GL4 12 as a function of temperature and redox potential. This procedure was independently repeated 5 times. Figure 5 contains the average values of E° for the 5 independent simulations, and the standard deviation in the E° predictions from the 5 independent simulations are shown with different colors. From simplicity, the full figure, that contains the same plots for the p$K_a$ of PRN 16 and GL4 12, is presented at Figure S1 in the Supporting Information. Results are shown for E,T,pH-REMD, E,T-REMD, E,pH-REMD and C(pH,E)MD.

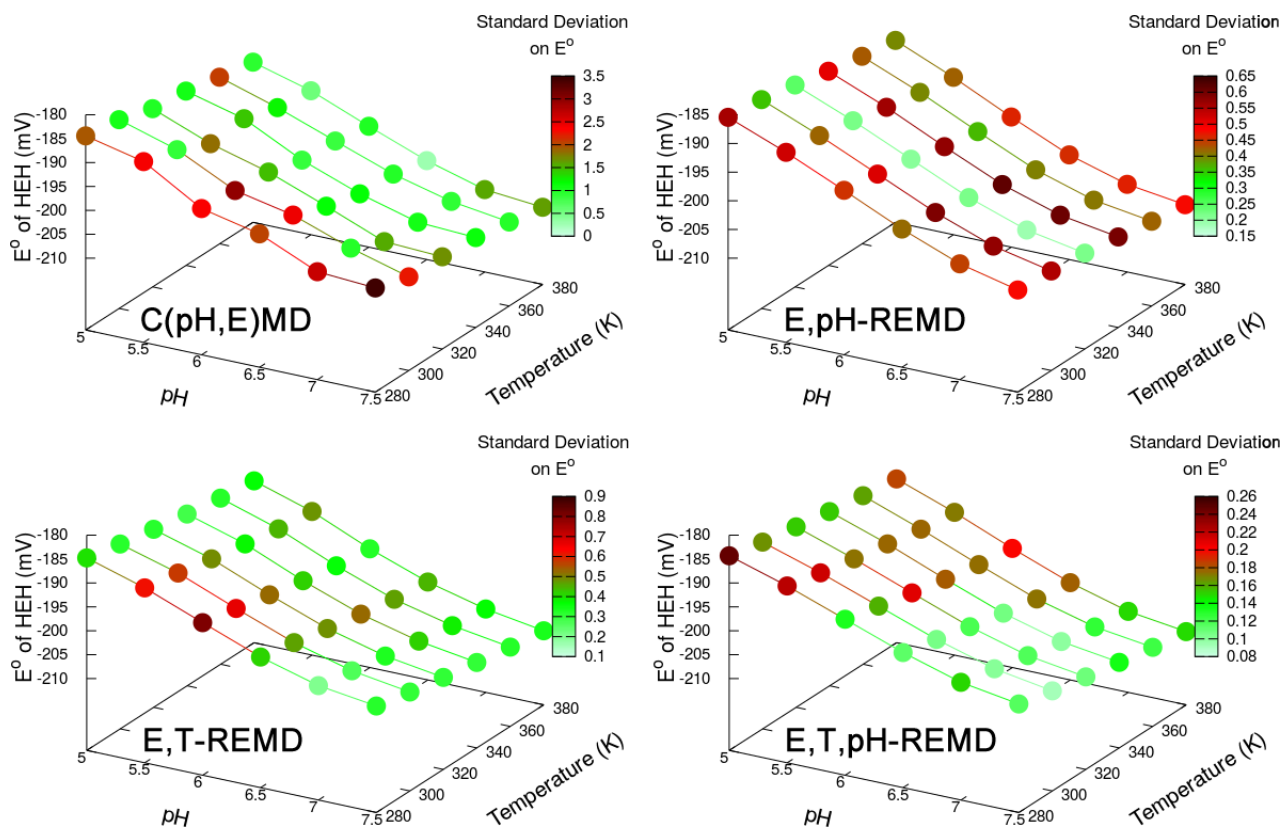

**Figure 5.** E° of HEH versus pH and temperature. The points represent average values for 5 independent simulations, and the colors, with their respective color bars, represent the standard deviations. Simulations are 5 ns longs.

C(pH,E)MD is the simulation without replica exchange attempts. In Figure 5 we notice that the REMD results are clearly more converged than the C(pH,E)MD results. This can be concluded not only based on the behavior of the average E° and p$K_a$ values (notice the C(pH,E)MD results for the p$K_a$ values in Figure S1), but also based on the standard deviations that are very significantly higher for C(pH,E)MD, as can be seen from the color bars.

In this analysis, we can also conclude that E,T-REMD yields more converged results than E,pH-REMD in the E° predictions. Looking at the E° predictions, we see that for E,pH-REMD most points have errors higher than 0.5 mV, while for E,T-REMD most points have errors below 0.4 mV. The comparison between E,T-REMD and E,pH-REMD for the p$K_a$ predictions is interesting because E,T-REMD does not contain exchanges in the pH dimension. As Figure S1 shows, E,pH-REMD gives smaller standard deviations than E,T-REMD for GL4 12, however, for PRN 16 it is E,T-REMD that gives smaller errors.

As Figure 5 shows, the addition of a third replica exchange dimension in E,T,pH-REMD gives more converged results than both E,T-REMD and E,pH-REMD. This is true for all residues, for both E° and p$K_a$ predictions. This can be quickly inferred just by looking at the maximum errors in the color bars. For example, for E° the maximum standard deviation in E,T,pH-REMD is 0.26 mV and this value is in the lower region of the color bars (light green colors) of both E,T-REMD and E,pH-REMD.

We also evaluated the effect of the order of dimensions being exchanged in three-dimensional REMD. Figure S2 in the Supporting Information contains results comparing E,T,pH-REMD, E,pH,T-REMD and pH,T,E-REMD, following the same procedures used in Figures 5 and S1. For both E° and p$K_a$ predictions, in Figure S2 we observe similar averages and standard deviations for all three-dimensional REMD simulations. Therefore, this indicates the order of dimensions being exchanged is not important in three-dimensional REMD simulations with pH, redox potential and temperature dimensions.

**Temperature dependence in predicting E° vs pH and p$K_a$ vs redox potential**

It has been shown, both experimentally [40,41] and theoretically in our previous publication [18], that E° should decrease when pH increases. Equivalently, p$K_a$ should also decrease when the redox potential raises. We now analyze the temperature dependence on E° and p$K_a$ predictions through a E,T,pH-REMD simulation. This simulation was executed for 54 ns considering a significantly higher number of replicas in comparison to the previous subsection. In this simulation, we considered 8 redox potential values from -293 to -83 mV in intervals of 30 mV, 6 target temperatures from 280 to 380 K in intervals of 20 K, and 12 pH values from 3.0 to 8.5 in intervals of 0.5, giving a total 576 replicas considered. Figure 6 shows the E° of HEH as a function of pH for each temperature, and the p$K_a$ of all pH-active residues and their sum as a function of the redox potential for each temperature. The solid lines are not fittings. They are the theoretical predictions from equations 7 and 8 and were obtained using the values of $E^o_{prot}$, $E^o_{deprot}$, $pK^{(i)}_{a,red}$ and $pK^{(i)}_{a,oxi}$ from the simulation data shown in the figure.

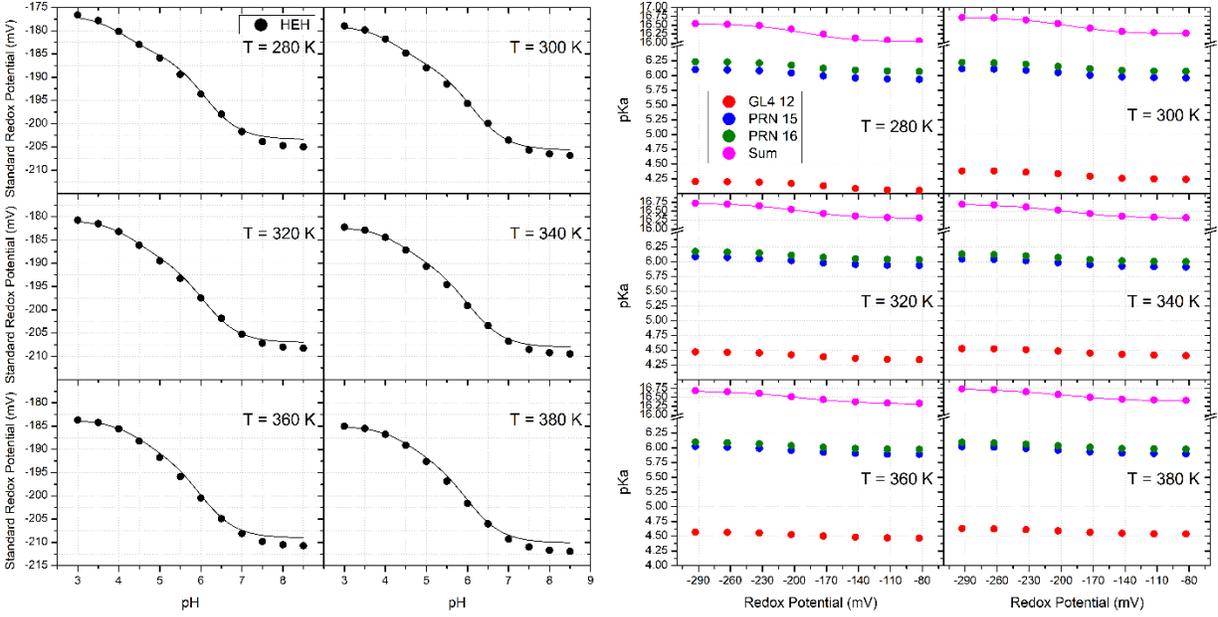

**Figure 6.** Standard redox potential (E°) of HEH as a function of pH, and p$K_a$ values of all pH-active residues as a function of the redox potential for different temperatures. Data obtained from a 54 ns long E,T,pH-REMD simulation. The solid lines were obtained using equations 7 and 8 with parameters from the simulation. GL4 = Glutamate, PRN = Propionate

From Figure 6 we observe that as the temperature raises the E° of HEH decreases. Further, the difference between $E^o_{prot}$ and $E^o_{deprot}$ decreases with increases in temperature. For the 6 temperatures from 280 to 380 K this difference is, respectively, 28.3, 27.8, 27.5, 27.3, 27.0, and 26.9 mV.

The pH-active residues have a different behavior. The p$K_a$ of the glutamate increases with temperate, and the difference between $pK_{a,red}^{(i)}$ and $pK_{a,oxi}^{(i)}$ for this residue decreases with temperature. For the glutamate the $pK_{a,red}^{(i)} - pK_{a,oxi}^{(i)}$ difference is 0.15 at 280 K and 0.09 at 380 K. Contrarily to the glutamate, for both propionates the p$K_a$ decreases with temperature, however, the difference between $pK_{a,red}^{(i)}$ and $pK_{a,oxi}^{(i)}$ for each propionate decreases when the temperature is raised, similarly to the glutamate. The $pK_{a,red}^{(i)} - pK_{a,oxi}^{(i)}$ difference is, for both propionates, 0.17 at 280 K and 0.12 at 380 K. Further, by comparing the p$K_a$ predictions for both propionates we see that the difference between their p$K_a$ values decreases with temperature. The difference between the p$K_a$ values of PRN 16 and PRN 15 is 0.13 at 280 K and 0.07 at 380 K.

By analyzing the solid lines in Figure 6 we conclude that a good description of the results can be made for all temperatures from the theoretical predictions in equations 7 and 8. This is achieved here without making any fittings. We also notice that E° versus pH and the p$K_a$ values versus redox potential at 300 K in Figure 6 match well with the results we obtained using E-REMD for different pH values at 300 K in our previous publication [18]. Also, in this previous publication we showed that our p$K_a$ predictions agree with experimental p$K_a$ values for other single heme proteins [18].

An important test we can do in our E,T,pH-REMD simulation is to check whether a given replica visits all the possible target redox potential, temperature and pH values. This is a test of both our methodology and of our choice for the distribution of target values. Figure 7 shows for 4 of our 576 replicas how long each replica spent during the E,T,pH-REMD simulation on each possible combination of redox potential, temperature and pH. Different counts for different target values are shown with different colors in Figure 7.

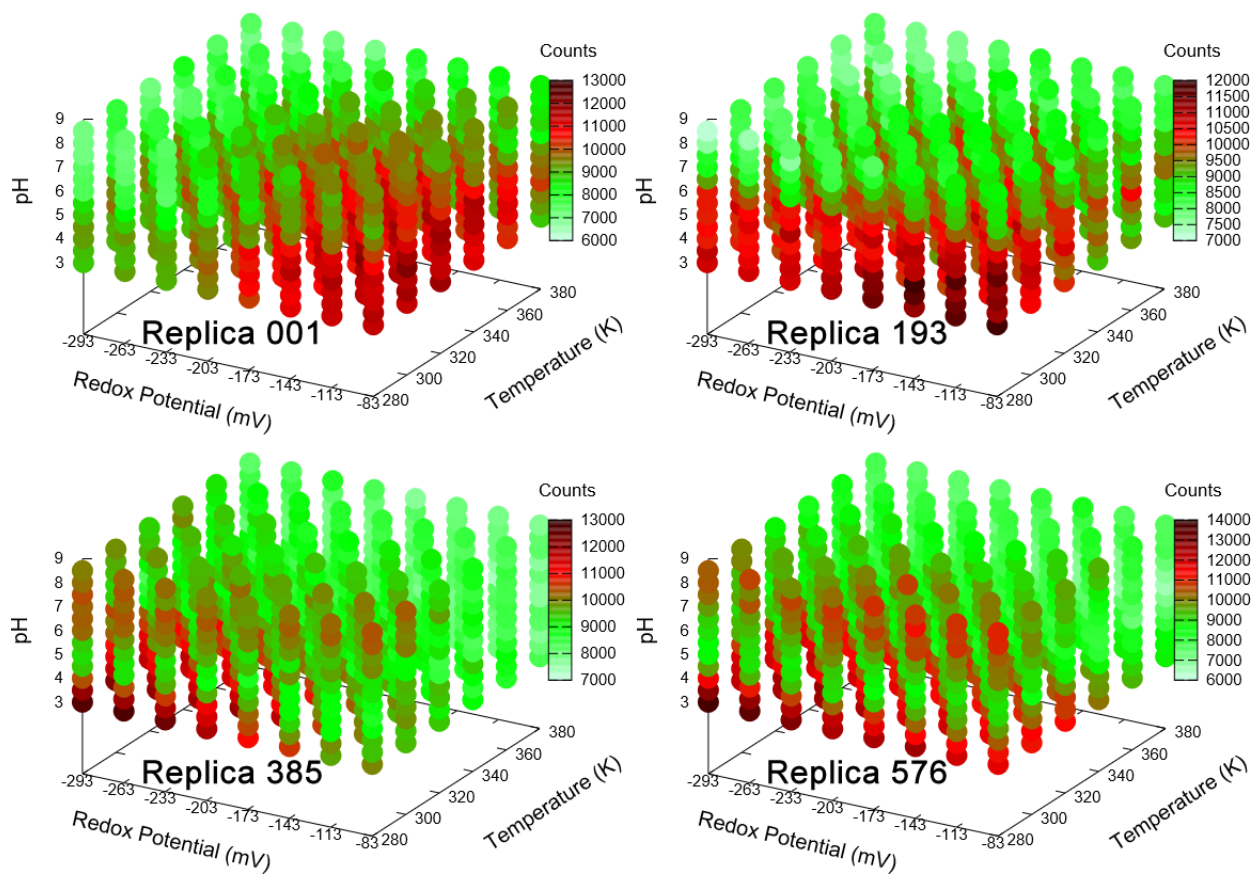

**Figure 7.** Distribution of target redox potential, temperature and pH values for 4 of the 576 replicas. The counts represent how long, during the E,T,pH-REMD simulation, each replica spent in a given combination of target values. Different counts are represented by different colors, as shown in the color bars.

As Figure 7 indicates, a given replica visits all possible combinations of target redox potential, temperature and pH during the E,T,pH-REMD simulation. As can be seen from the color bars, the counts for the most populated combinations of target values are at most 2.2 times larger than the counts for the least populated combinations. This shows our choice for the distributions of target values is good and yields satisfactory exchange acceptance rates. Also, as one would expect, we see that each replica populates the target values differently. As the figure shows, the most populated target values are distributed differently for the different replicas.

**Computational Benchmarks**

In Table 1 we compare the computational performance of the AMBER module *pmemd* for different computations. Calculations were performed at the Blue Waters supercomputer using Cray XK7 nodes. Each node contains a Tesla K20X GPU. Our system, composed by NAcMP8 and the histidine peptide, has 237 atoms. We performed calculations for 216 [pH,E,T] values by combining $pHs = 5.0, 5.5, 6.0, 6.5, 7.0$ and $7.5$, $Es = -263, -233, -203, -173, -143$ and $-113$ mV, and $Ts = 280, 300, 320, 340, 360$ and $380$ K. The computational performances shown in Table 1 are the averages for the 216 [pH,E,T] replicas.

| Computation | pmemd.MPI (2 CPUs/rep.) | pmemd.MPI (16 CPUs/rep.) | pmemd.cuda.MPI (1 GPU/rep.) |
|---|---|---|---|
| | Computational performance (ns/day) | | |
| Regular MD | 43 | 216 | 400 |
| C(pH,E)MD | 24 | 134 | 237 |
| E-REMD | 21 | 99 | 186 |
| pH-REMD | 21 | 98 | 183 |
| T-REMD | 22 | 98 | 176 |
| E,pH-REMD | 21 | 103 | 185 |
| E,T-REMD | 20 | 98 | 180 |
| E,pH,T-REMD | 21 | 100 | 181 |

**Table 1.** Computational performances of *pmemd* for different computations. Performances shown are averages for 216 replicas.

As Table 1 shows, the multidimensional REMD simulations have essentially the same computational performance as the one-dimensional simulations. The REMD simulations are approximately 55% slower than regular MD, and 23% slower than C(pH,E)MD (limit of EAF = 0). Regarding the GPU-accelerated CUDA calculations, we see that speedups are obtained: the REMD calculations are 8.7 times faster than using 2 CPUs and 1.8 times faster than using 16 CPUs.

It is important to mention that, as we showed in our previous publication [18], the computational performances are different in explicit solvent where the number of atoms is significantly higher. In explicit solvent the GPU-accelerated calculations become much more efficient in comparison to CPUs.

CONCLUSIONS

We presented the implementation of pH and redox potential as options for multidimensional REMD simulations in AMBER. These implementations are available using both implicit and explicit solvent models, however, for simplicity only implicit solvent results were shown in this work. Previous publications have shown that the addition of replica exchange significantly improves statistical convergence in comparison to simulations without replica exchange [3,18,19]. In this work we have shown that the addition of more replica exchange dimensions further improves sampling convergence on constant pH and redox potential simulations in comparison to one-dimensional REMD results: we observed that E,pH-REMD yields more converged results than both E-REMD and pH-REMD, that E,T-REMD yields more converged results than both E-REMD and T-REMD, and also that E,T,pH-REMD yields more converged results than both E,pH-REMD and E,T-REMD. These findings corroborate with previous multidimensional REMD publications that showed H,T-REMD performs better than H-REMD and T-REMD [25,26].

An important aspect to evaluate is: what is the best dimension to be added to a given one-dimensional REMD simulation in order to obtain a better convergence efficiency? In this work we addressed this question by comparing the addition of pH or T dimensions to E-REMD. This was done by analyzing the data at 300 K and pH 7.0. This is the only combination of target temperature and pH values that exists in both our E,pH-REMD and E,T-REMD simulations. We conclude that E,T-REMD yields more converged E° predictions than E,pH-REMD. Even though, the application of E,pH-REMD may become handy in comparison to E,T-REMD in situations where the number of replicas required for a good exchange rate in the temperature dimension is high. This is true when the number of atoms in the system is large, as is generally the case in explicit solvent simulations.

Through a E,T,pH-REMD simulation we were able to observe for our system, NAcMP8 with a histidine peptide, the behavior of the E° of HEH versus pH and temperature, and the behavior of the p$K_a$ of all pH-active residues as a function of redox potential and temperature. As one would expect, in our simulations E° decreases when pH increases and the p$K_a$s decrease as the redox potential increases. The temperature effect varies for the different residues: the E° of HEH decreases with temperature, the p$K_a$ of the glutamate increases with temperature, and the p$K_a$ of both propionates decrease with temperature. Also, all the results in our simulation are shown to be in good agreement with theoretical predictions devised for the behavior of E° vs pH and T and of $\sum_i pK_a^{(i)}$ vs redox potential and T. We also verified that a given replica in our E,T,pH-REMD simulation visits all possible combinations of target redox potential, temperature and pH values throughout the simulation. This validates our multidimensional REMD approach of performing sequential one-dimensional exchange attempts and indicates our choice of target values produced satisfactory rates of acceptance for the exchanges.

Perhaps the most important conclusion in this work summarizes as: the more dimensions we add in a multidimensional REMD simulation, the more efficiently we reach ensemble convergence. In terms of computational performance, adding more replica exchange dimensions is not an issue because, as our computational benchmarks show, the computational performance of multidimensional REMD with redox potential, temperature and pH dimensions is essentially the same as of the one-dimensional REMD simulations. However, more computational resources are required as the number of replicas significantly increases in comparison to one-dimensional REMD. In multidimensional REMD the number of replicas geometrically increases with the number of dimensions. It is important to mention that our computational benchmarks have shown that our multidimensional REMD implementations have a bearable computational performance in comparison to regular MD. Also, high computational performance is obtained with AMBER's GPU-accelerated code in comparison to calculations using only CPUs.

AUTHOR INFORMATION

**Corresponding Author**

* E-mail: roitberg@ufl.edu

**Notes**

The authors declare no competing financial interest.

ACKNOWLEDGMENTS

The authors gratefully acknowledge financial support from CAPES (Brazil). This research is part of the Blue Waters sustained-petascale computing project, which is supported by the National Science Foundation (awards OCI-0725070 and ACI-1238993) and the state of Illinois. Blue Waters is a joint effort of the University of Illinois at Urbana-Champaign and its National Center for Supercomputing Applications.